\documentclass[10pt]{article}

\usepackage[super,compress]{cite}

\usepackage{fancyhdr}
\usepackage{amsmath}
\usepackage{amsfonts}
\usepackage{latexsym}
\usepackage{epsfig}
\usepackage{graphicx}
\usepackage{verbatim}
\usepackage{psfrag}
\usepackage{multirow}
\usepackage{ragged2e}
\usepackage{cancel}
\usepackage{color}
\usepackage{xcolor}
\usepackage{hyperref}

\newcommand{\ba}{\begin{eqnarray}}
\newcommand{\ea}{\end{eqnarray}}
\newcommand{\be}{\begin{equation}}
\newcommand{\ee}{\end{equation}}
\newcommand{\bd}{\begin{displaymath}}
\newcommand{\ed}{\end{displaymath}}
\newcommand{\bn}{\begin{enumerate}}
\newcommand{\en}{\end{enumerate}}

\newcommand{\bp}{\begin{pmatrix}}
\newcommand{\ep}{\end{pmatrix}}

\definecolor{darkgreen}{rgb}{0.,0.6,0.2}

\definecolor{darkgreen}{rgb}{0.,0.6,0.2}

\date{}

\begin{document}

\title{Re-evaluation of Spin-Orbit Dynamics of Polarized $e^+ e^-$ Beams in High Energy
Circular Accelerators and Storage Rings: an approach based on a 
Bloch equation{\thanks{
Based on a talk at IAS, Hong Kong, January 17, 2019. Also
available as article:
\emph{Int. J. Mod. Phys.}, vol.\,A35, Nos.\,15 {\&}\,16, 2041003, 2020.
Moreover available as DESY Report 20-137.
}}}

\maketitle



\centerline{Klaus Heinemann \footnote{Corresponding author.} }
\centerline{\em Department of Mathematics and Statistics, University of New Mexico,} 
\centerline{\em Albuquerque, NM 87131, USA} 
\centerline{\em heineman@math.unm.edu}   
\vspace{4mm}
\centerline{Daniel Appel\"o \footnote{Now at Michigan State University, USA. (email:  appeloda@msu.edu)} } 
\centerline{\em Department of Applied Mathematics, University of Colorado Boulder,} 
\centerline{\em Boulder, CO 80309-0526, USA } 
\centerline{\em Daniel.Appelo@Colorado.edu}  
\vspace{4mm}
\centerline{\em Desmond P. Barber} 
\centerline{\em Deutsches Elektronen-Synchrotron (DESY) }
\centerline{\em Hamburg, 22607, Germany}
\centerline{\em and:}
\centerline{\em Department of Mathematics and Statistics, University of New Mexico}
\centerline{\em Albuquerque, NM 87131, USA}
\centerline{\em mpybar@mail.desy.de}

\vspace{4mm}
\centerline{\em Oleksii Beznosov}
\centerline{\em Department of Mathematics and Statistics, University of New Mexico,} 
\centerline{\em Albuquerque, NM 87131, USA}
\centerline{\em dohter@protonmail.com}

\vspace{4mm}
\centerline{\em James A. Ellison}
\centerline{\em Department of Mathematics and Statistics, University of New Mexico,} 
\centerline{\em Albuquerque, NM 87131, USA}
\centerline{\em ellison@math.unm.edu}

\maketitle


\newpage

\begin{abstract}

We give an overview of our current/future analytical and
numerical work on the spin polarization in 
high-energy electron storage rings. Our goal is to study the 
possibility of polarization for the CEPC and FCC-ee.
Our work is based on the so-called 
Bloch equation for the polarization density introduced by
Derbenev and Kondratenko in 1975.
We also give an outline of the standard approach, the
latter being based on the Derbenev-Kondratenko formulas.

\vspace{5mm}

{\tiny{.}} \hspace{12mm} differential equations, method of averaging. 
\end{abstract}



\newpage

\tableofcontents

\newpage

\section{Introduction}
This paper is an update on a talk by K. Heinemann
at the IAS Mini-Workshop on
Beam Polarization in Future Colliders on January 17, 2019, in Hong Kong
\cite{IAS}.
Our ultimate goal is to examine the 
possibility of high polarization for CEPC and FCC-ee.

We will first briefly review the ``standard'' approach
which is based on the Derbenev-Kondratenko formulas \cite{DK73}.
These formulas rely, in part, on plausible assumptions 
grounded in deep physical intuition. So the following question arises:
do the Derbenev-Kondratenko formulas tell full story? In fact there is an
alternative approach based on a Bloch-type equation for the 
polarization density \cite{DK75}
which we call
\footnote{Note that in previous work we sometimes 
called it the full Bloch equation.}
the Bloch equation (BE)
and which we believe can deliver more information 
than the standard approach 
even if the latter includes potential correction terms \cite{DBBQ}.
So we aim to determine 
the domain of applicability of the Derbenev-Kondratenko formulas
and the possibility in theory of polarization at the CEPC and FCC-ee
energies.
Of course both approaches focus on the equilibrium polarization and the
polarization time.
We use the name ``Bloch'' to reflect the analogy with equations
for magnetization in condensed matter \cite{Bloch}.
This paper concentrates on the Bloch approach. 
The cost of the numerical computations in the Bloch approach is 
considerable since the polarization density depends on six phase-space
variables plus the time variable so that the numerical 
solution of the BE, the BE being a system of three
PDEs in seven independent variables, is a nontrivial task which cannot
be pursued with traditional approaches like the finite difference method.
However we see at least five viable methods:
\begin{enumerate} 
\item Approximating the BE by an effective 
BE via the Method of Averaging and solving the 
effective BE via spectral phase-space discretization, e.g.,
a collocation method, plus an implicit-explicit
time discretization.
\item Solving the system of stochastic differential 
equations (SDEs), which underlies the
BE, via Monte-Carlo spin tracking. See Ref. \citen{SDE} for
the system of SDEs underlying the BE.
\item Solving the Fokker-Planck equation, which underlies the
BE, via the Gram-Charlier method.
\item Solving the BE via a deep learning method.
\item Solving the system of SDEs
in a way that allows connections 
with the Derbenev-Kondratenko formulas to be established.
\end{enumerate}

We will dwell on Method 1 in this paper. 
We plan to validate this method
by one of the other four methods. 
More details on Method 1 can be found in Ref. \citen{SDE}.
The method of averaging we use is discussed in 
Refs. \citen{El}-\citen{CE}.
One hope tied to Method 1 is that the effective BE gives
analytical insights into the spin-resonance structure of the bunch.
Note that Methods 1-4 are independent of the standard approach. In
particular they do not rely on the invariant spin field.
Note also that Methods 1-3 and 5 are based on knowing the 
system of SDEs, which underlies the BE.
For details of this system of SDEs, see the invited ICAP18 paper
of Ref. \citen{SDE}. Regarding Method 2 there is a large literature
on the numerical solution of SDEs, see Refs. \citen{KP},
\citen{KPS} and references in Ref. \citen{PB}.

By neglecting
the spin-flip terms and the kinetic-polarization term in the BE
one obtains an equation that we call the
Reduced Bloch equation (RBE).
The RBE approximation is sufficient for computing the 
radiative depolarization rate due to stochastic orbital effects
and it shares the terms with the BE that are challenging to discretize. For details on our phase-space discretization and 
time discretization of the RBE, 
see Refs. \citen{SDE},\citen{thesis},\citen{HABBE} and \citen{OB}. 

We proceed as follows. In Section 2 we
sketch the standard approach. In Section 3 we present, for the
laboratory frame, the BE and its restriction, the RBE.
In Section 4 we discuss the RBE in the beam frame and
in Section 5 we show how, in the beam frame,
the effective RBE is obtained via
the method of averaging. In Section 6 we describe ongoing and future work.
\section{Sketching the standard approach based on the
Derbenev-Kondratenko formulas}
We define the ``time'' $\theta = 2\pi s/C$ where $s$ is the distance around the ring and $C$ is the circumference. We denote by $y$ a position
in six-dimensional phase space of accelerator coordinates which
we call beam-frame coordinates. 
In particular, following Ref. \citen{BR},
$y_6$ is the relative deviation of the energy from the reference  energy.
Then if, 
$f=f(\theta,y)$ denotes the normalized $2\pi$-periodic 
equilibrium phase-space density
at $\theta$  and $y$ and  
$\vec{P}_{loc}=\vec{P}_{loc}(\theta,y)$ denotes the local polarization 
vector of the bunch we have
\begin{eqnarray}
&& \int dy\; f(\theta,y) = 1 \; , \quad 
 \int dy\; f(\theta,y) \vec{P}_{loc}(\theta,y) =
\vec{P}(\theta) \; ,
\label{eq:2.10}
\end{eqnarray}
where $\vec{P}(\theta)$ is the polarization vector of the bunch at 
$\theta$.
For a detailed discussion about  $\vec{P}_{loc}$, see, e.g., Ref. \citen{BH}.
Here and in the following we use arrows on three-component column vectors.

Central to the standard approach is the 
invariant spin field (ISF) $\hat{n}=\hat{n}(\theta,y)$
defined as a normalized periodic solution of
the Thomas-BMT-equation in phase space, i.e.,
\ba
&& \partial_\theta\hat{n}
= L_{\rm Liou}(\theta,y)\hat{n}   
+ \Omega(\theta,y)\hat{n} \; , 
\label{eq:2.20} 
\ea
such that
\begin{enumerate} 
\item $\Big{|}\hat{n}(\theta,y)\Big{|}=1$,
\item $\hat{n}(\theta+2\pi,y)=\hat{n}(\theta,y)$,
\end{enumerate}
and where $L_{\rm Liou}$ denotes the Hamiltonian part of the  
Fokker-Planck operator $L_{\rm FP}^y$, the latter being introduced in
Section 3 below. For some of our work on the ISF see Refs. \citen{EH}
and \citen{Spin1B}.
The unit vector of the ISF on the closed orbit is denoted by $\hat n_0(\theta)$ and it
is easily obtained as an eigenvector of the one-turn spin-transport map on the 
closed orbit \cite{BR}.
There are many methods for computing the ISF but
none are trivial 
(for a recent technique see Ref. \citen{Sag2}).
In fact the 
existence, in general, of the invariant spin field is a mathematical issue
which is only partially resolved, see, e.g., Ref. \citen{EH}.
The standard approach assumes that a function
$P_{\rm DK}=P_{\rm DK}(\theta)$ exists such that
\begin{eqnarray} 
&& \vec{P}_{loc}(\theta,y)\approx P_{\rm DK}(\theta)\hat{n}(\theta,y) \; .
\label{eq:2.13}
\end{eqnarray}
Thus, by (\ref{eq:2.10}) and (\ref{eq:2.13}),
\begin{eqnarray}
&& \vec{P}(\theta) = \int dy\; f(\theta,y) \vec{P}_{loc}(\theta,y)
\approx P_{\rm DK}(\theta) \int dy\; f(\theta,y)\hat{n}(\theta,y)
\; .
\label{eq:2.21}
\end{eqnarray}

The approximation (\ref{eq:2.13}) leads to \cite{DK73}
\begin{eqnarray} 
&& P_{\rm DK}(\theta) 
= P_{\rm DK}(\infty)(1 - e^{-\theta/\tau_{\rm DK}}) 
+ P_{\rm DK}(0)e^{-\theta/\tau_{\rm DK}} \; ,
\label{eq:2.25}
\end{eqnarray}
where $\tau_{\rm DK}$ and $P_{\rm DK}(\infty)$ are given by 
the Derbenev-Kondratenko formulas
\begin{eqnarray} 
&& \hspace{-15mm}
P_{\rm DK}(\infty):= \frac{\tau_0^{-1}}{\tau_{\rm DK}^{-1}} \; ,
\label{eq:2.30}\\
&& \hspace{-15mm} \tau_{\rm DK}^{-1}:= 
\frac{5\sqrt{3}}{8}\frac{r_e\gamma_0^5\hbar}{m}
\frac{C}{4\pi^2}\int_0^{2\pi}\;d\theta 
\Big{\langle} \frac{1}{|R|^3}
[1-\frac{2}{9}(\hat{n}\cdot{\hat{\beta}})^2+\frac{11}{18}
\Big{|}\partial_{y_6}
\hat{n}\Big{|}^2]
\Big{\rangle}_\theta \; ,
\label{eq:2.31} \\
&& \hspace{-15mm}
\tau_0^{-1} := \frac{r_e\gamma_0^5\hbar}{m}
\frac{C}{4\pi^2}\int_0^{2\pi}\;d\theta 
\Big{\langle}
\frac{1}{|R|^3}
\hat{b}\cdot[\hat{n}-\partial_{y_6}\hat{n}]\Big{\rangle}_\theta \; ,
\label{eq:2.31new}
\end{eqnarray}
with 
\begin{itemize}
\item $\Big{\langle}\cdots\Big{\rangle}_\theta\equiv\int dy\; f(\theta,y)\cdots$
\item  
$\hat{b}=\hat{b}(\theta,y)
\equiv$ normalized magnetic field,
$\hat{\beta}=\hat{\beta}(\theta,y)
\equiv$ normalized velocity vector,
$\gamma_0\equiv$ Lorentz factor of the reference particle,
$R(\theta,y)\equiv$ radius of curvature in the  external magnetic field,
$r_e\equiv$ classical electron radius, $m \equiv$ rest mass of electrons
or positrons.
\end{itemize}
By (\ref{eq:2.21}) and for large $\theta$
\begin{eqnarray} 
&&  \vec{P}(\theta) \approx P_{\rm DK}(\infty)
\int dy\; f(\theta,y)\hat{n}(\theta,y) \; ,
\label{eq:2.32new} 
\end{eqnarray}
where $P_{\rm DK}(\infty)$ is given by (\ref{eq:2.30}) and
where the rhs of (\ref {eq:2.32new}) is the approximate equilibrium polarization vector.
Note that the latter is $2\pi$-periodic in $\theta$ since 
$f(\theta,y)$ and $\hat{n}(\theta,y)$ are $2\pi$-periodic in $\theta$.
Defining
\begin{eqnarray} 
&& \tau_{dep}^{-1}:=\frac{5\sqrt{3}}{8}\frac{r_e\gamma_0^5\hbar}{m}
\frac{C}{4\pi^2}\int_0^{2\pi}\;d\theta 
\Big{\langle}
\frac{1}{|R|^3}
\frac{11}{18}
\Big{|} \partial_{y_6}\hat{n}\Big{|}^2
\Big{\rangle}_\theta \; ,
\label{eq:2.32} 
\end{eqnarray}
we can write (\ref{eq:2.31}) as
\begin{eqnarray} 
&& \tau_{\rm DK}^{-1}= \tau_{dep}^{-1}+
\frac{5\sqrt{3}}{8}\frac{r_e\gamma_0^5\hbar}{m}
\frac{C}{4\pi^2}\int_0^{2\pi}\;d\theta 
\Big{\langle} \frac{1}{|R|^3}
[1-\frac{2}{9}(\hat{n}\cdot{\hat{\beta}})^2]\Big{\rangle}_\theta \; .
\label{eq:2.33} 
\end{eqnarray}
For details on (\ref{eq:2.30}), (\ref{eq:2.31}), (\ref{eq:2.31new}),
(\ref{eq:2.32}) and (\ref{eq:2.33}) 
see, e.g., Refs. \citen{MSY} and \citen{BR}.

We now briefly characterize the various terms in 
the Derbenev-Kondratenko formulas.
First, $\tau_{dep}^{-1}$ is the radiative depolarization rate. Secondly,
the term $\frac{r_e\gamma_0^5\hbar}{m}
\frac{C}{4\pi^2}\int_0^{2\pi}\;d\theta 
\Big{\langle}
\frac{1}{|R|^3}
\hat{b}\cdot\hat{n}\Big{\rangle}_\theta$ 
in $\tau_0^{-1}$ and the term
$\frac{5\sqrt{3}}{8}\frac{r_e\gamma_0^5\hbar}{m}
\frac{C}{4\pi^2}\int_0^{2\pi}\;d\theta 
\Big{\langle} \frac{1}{|R|^3}
\Big{\rangle}_\theta$ in $\tau_{\rm DK}^{-1}$ cover the Sokolov-Ternov 
effect. Lastly, the term
$-\frac{r_e\gamma_0^5\hbar}{m}
\frac{C}{4\pi^2}\int_0^{2\pi}\;d\theta 
\Big{\langle}
\frac{1}{|R|^3}
    \hat{b}\cdot[\partial_{y_6}\hat{n}]
\Big{\rangle}_\theta$ in $\tau_0^{-1}$ covers the kinetic 
polarization effect and the term in
$\tau_{\rm DK}^{-1}$ which is proportional to $2/9$ covers the
Baier-Katkov correction. 

We now sketch three approaches for computing $P_{\rm DK}(\infty)$ via the 
Derbenev-Kondratenko formulas.
All three approaches use (\ref{eq:2.30}) but they differ in how
$\tau_0^{-1}$ and $\tau_{\rm DK}^{-1}$ are computed.
\begin{itemize}
\item[(i)] Compute $\tau_0^{-1}$ via (\ref{eq:2.31new}) and $\tau_{\rm DK}^{-1}$
via (\ref{eq:2.31}) by computing $f$ and $\hat{n}$ as accurately as needed.
\item[(ii)]
Approximate  $\tau_0^{-1}$ by neglecting the usually-small kinetic 
polarization term in
(\ref{eq:2.31new}) and by approximating the remaining term in
(\ref{eq:2.31new}) by replacing $\hat{n}$ by $\hat{n}_0$.
Compute $\tau_{\rm DK}^{-1}$ via (\ref{eq:2.33}) 
where $\tau_{dep}^{-1}$ is not computed via (\ref{eq:2.32}) 
but via Monte-Carlo spin tracking and where 
the remaining terms in (\ref{eq:2.33}) are approximated
by using the $\hat{n}_0$-axis. 
\footnote{Prominent Monte-Carlo spin tracking codes are 
SLICKTRACK by D.P. Barber \cite{BR}, SITROS by J. Kewisch \cite{BR},
Zgoubi by F. Meot \cite{Zg},
PTC/FPP by E. Forest \cite{Fo}, and Bmad by D. Sagan \cite{Sag}.
This approach provides 
a useful first impression avoiding the computation of 
$f$ and $\hat{n}$. For more details on this approach see Ref.
\citen{BR}. Monte-Carlo tracking can also be extended
beyond integrable orbital motion to include, as just one example,
beam-beam forces.
Note that Monte-Carlo tracking just gives an estimate 
of  $\tau_{dep}^{-1}$ but it does not provide an explanation.
Nevertheless, insights into sources of depolarization can be
obtained by switching off terms in the Thomas-BMT equation. 
In principal such diagnoses can also be applied in approach (i).
Such investigations can the systematized under the heading of
``spin matching'' \cite{BR}.}
\item[(iii)] Compute $\tau_0^{-1}$ via (\ref{eq:2.31new}) and $\tau_{\rm DK}^{-1}$
via (\ref{eq:2.31}) by linear approximation in orbit and spin variables via
the so-called SLIM formalism \cite{BR}.
\end{itemize}
Approach (ii) is the most practiced while approach (i) is only
feasible if one can compute $f$ and $\hat{n}$ as accurately as needed
(which is not easy!). Approach (iii), which was historically the first,
is very simple and is often used for ballparking $P_{\rm DK}(\infty)$.
Since the inception of the Derbenev-Kondratenko formulas
correction terms to the rhs of (\ref{eq:2.32}) have been suspected.
See Refs. \citen{DBBQ},
\citen{DK72} as well as Z. Duan's contribution to this
workshop. These correction terms, associated with so-called 
resonance crossing, in turn associated with large energy spread,  are not as well understood as the rhs
of (\ref{eq:2.32}), partly because of their peculiar form.
Nevertheless, careful observation of spin motion during the Monte-Carlo 
tracking in approach (ii),
might provide a way to investigate their existence and form.

\section{The Bloch equation and the Reduced
Bloch equation in the laboratory frame}

In the previous section
we used the beam frame and we will do so later.
However the BE was first presented in Ref. \citen{DK75} 
for the laboratory frame and in that frame it also has its simplest form. 
In this section we focus on the laboratory frame.

In a semiclassical probabilistic description of an electron 
or positron bunch the
spin-orbit dynamics is described by the
{\it spin-$1/2$ Wigner function} $\rho$ 
(also called the {\it Stratonovich function})
written as
\begin{eqnarray}
 \rho(t,z) = \frac{1}{2} [ f_{lab}(t,z) I_{2\times 2}
+ \vec{\sigma}\cdot\vec{\eta}_{lab}(t,z) ]  \; ,
               \label{eq:rho}
\end{eqnarray}
with $z=(\vec{r},\vec{p})$ where $\vec{r}$ and $\vec{p}$
are the position and momentum vectors
of the phase space and $t$ is the time, and
where $f_{lab}$ is the phase-space density of particles
normalized by $\int dz f_{lab}(t,z) = 1$,
$\vec{\eta}_{lab}$ is the polarization density of the bunch and
$\vec{\sigma}$ is the vector of the three Pauli matrices.
As explained in Ref. \citen{BH}, $\vec{\eta}_{lab}$ is proportional to the
spin angular momentum density.
In fact it is given by $\vec{\eta}_{lab}(t,z)=f_{lab}(t,z)\vec{P}_{loc,lab}(t,z)$
where $\vec{P}_{loc,lab}$ is the
local polarization vector.
Thus $f_{lab}=Tr[\rho]$ and
$\vec{\eta}_{lab}=Tr[\rho\vec{\sigma}]$.
The polarization vector $\vec{P}_{lab}(t)$ of the bunch is
$\vec{P}_{lab}(t) = \int dz\vec{\eta}_{lab}(t,z)
 = \int dz f_{lab}(t,z)\vec{P}_{loc,lab}(t,z)$.

Then, by neglecting collective effects and after several other approximations,
the phase-space density evolves according to Ref. \citen{DK75}
via
\ba
&& \partial_t f_{lab}  =  L_{FP}^{lab}(t,z)f_{lab} \; .
\label{eq:n10a}
\ea
Using the units as in Ref. \citen{DK75}
the Fokker-Planck operator $L_{FP}^{lab}$ is defined by
\ba
&&
\hspace{-10mm}
L_{FP}^{lab}(t,z) := L_{Liou}^{lab}(t,z)
+\vec{F}_{rad}(t,z)+\vec{Q}_{rad}(t,z)
+\frac{1}{2}\sum_{i,j=1}^3 \partial_{p_i}\partial_{p_j}
{\cal E}_{ij}(t,z) \; ,
\label{eq:4.4.4}
\ea
where
%
\ba
&& L_{Liou}^{lab}(t,z):= 
-\partial_{\vec{r}}\cdot\frac{1}{m\gamma(\vec{p})}\vec{p}
-\partial_{\vec{p}}\cdot [e\vec{E}(t,\vec{r})+\frac{e}{m\gamma(\vec{p})}(\vec{p}
\times\vec{B}(t,\vec{r}))] \; , 
\label{eq:4.4.4new}\\
&&
\vec{F}_{rad}(t,z):
=-\frac{2}{3} \frac{e^4}{m^5\gamma(\vec{p})}
|\vec{p}\times\vec{B}(t,\vec{r})|^2
\vec{p}\; ,
\label{eq:4.4.9b} \\
&& \vec{Q}_{rad}(t,z):=\frac{55}{48\sqrt{3}}
\sum_{j=1}^3\;
\frac{\partial[\lambda(t,z)\vec{p} p_j]}{\partial p_j} \; ,
\label{eq:4.4.11} \\
&&
{\cal E}_{ij}(t,z):=\frac{55}{24\sqrt{3}}\lambda(t,z)p_i p_j  \; , \quad
\lambda(t,z)
:= \hbar \frac{|e|^5}{m^8\gamma(\vec{p})}
|\vec{p}\times\vec{B}(t,\vec{r})|^3 \; ,
\label{eq:4.4.9c} \\
&&
\gamma(\vec{p}):=\frac{1}{m}\sqrt{|\vec{p}|^2+m^2} \; ,
\label{eq:4.4.9d}
\end{eqnarray}
and with $e$ being the electric charge 
of the electron or positron
and $\vec{E}$ and $\vec{B}$
being the external electric and magnetic fields. 

The Fokker-Planck operator $L_{FP}^{lab}$ whose explicit form is taken from 
Ref. \citen{DK75}
is a linear second-order partial
differential operator and, with some additional approximations, is
commonly used for electron synchrotrons and
storage rings, see Ref. \citen{Sa} and
Section 2.5.4 in Ref. \citen{BR}.
As usual, since it is minuscule compared to all other forces,
the Stern-Gerlach effect from the spin onto the
orbit is neglected in (\ref{eq:n10a}).
The polarization density $\vec{\eta}_{lab}$ evolves via eq. 2 in 
Ref. \citen{DK75}, i.e., via that which we call the 
Bloch equation,
namely
\ba
&& \partial_t\vec{\eta}_{lab}  =
L_{\rm FP}^{lab}(t,z)\vec{\eta}_{lab}+ M(t,z)\vec{\eta}_{lab}
\nonumber\\
&&\quad
-[1+\partial_{\vec{p}}\cdot \vec{p}]
\lambda(t,z)\frac{1}{m\gamma(\vec{p})}
\frac{\vec{p}\times
\vec{a}(t,z)}{|\vec{a}(t,z)|}f_{lab}(t,z) \; ,
%
\label{eq:n10}
\ea
where
\ba
&& M(t,z):=\Omega^{lab}(t,z)-\lambda(t,z)\frac{5\sqrt{3}}{8}[I_{3\times 3}
-\frac{2}{9m^2\gamma^2(\vec{p})}\vec{p}\vec{p}^T] \; ,
\label{eq:4.4.17} \\
&& \vec{a}(t,z):=\frac{e}{m^2\gamma^2(\vec{p})}(\vec{p}\times
\vec{B}(t,\vec{r})) \; .
\label{eq:4.4.17a} 
\ea
The BE was derived in Ref. \citen{DK75} 
from the semiclassical approximation 
of quantum electrodynamics and it is a generalization, 
to the whole phase space,
of the Baier-Katkov-Strakhovenko equation which just describes the
evolution of polarization along a single deterministic trajectory \cite{BKS}.
Note also that, while the BE was new in 1975, 
the orbital Fokker-Planck equation (\ref{eq:n10a}) 
was already known thanks to
research of the 1950s, e.g., Schwinger's paper on quantum corrections
to synchrotron radiation \cite{Sch}. 
The skew-symmetric matrix $\Omega^{lab}(t,z)$
takes into account the Thomas-BMT spin-precession effect.
Thus in the laboratory frame the
Thomas-BMT-equation (\ref{eq:2.20}) reads as
%
\ba
&& \partial_t\hat{n}_{lab}
= L_{Liou}^{lab}(t,z)
\hat{n}_{lab}  
+ \Omega^{lab}(t,z)\hat{n}_{lab} \; .
\ea
The quantum aspect of 
(\ref{eq:n10a}) and (\ref{eq:n10}) 
is embodied in the factor 
$\hbar$ in $\lambda(t,z)$. For example $\vec{Q}_{rad}$ is
a quantum correction to the classical radiation reaction force $\vec{F}_{rad}$.
~The ~terms $-\lambda(t,z)\frac{5\sqrt{3}}{8}\vec{\eta}_{lab}$ 
~and ~~$-\lambda(t,z)\frac{1}{m\gamma(\vec{p})}
\frac{\vec{p}\times\vec{a}(t,z)}{|\vec{a}(t,z)|}f_{lab}(t,z)$
take into account spin flips due to synchrotron radiation
and encapsulate the Sokolov-Ternov effect.
The term $\lambda(t,z)\frac{5\sqrt{3}}{8}
\frac{2}{9m^2\gamma^2(\vec{p})}\vec{p}\vec{p}^T\vec{\eta}_{lab}$
encapsulates the Baier-Katkov correction, and the term
$\partial_{\vec{p}}\cdot \vec{p}
\;\lambda(t,z)\frac{1}{m\gamma(\vec{p})}
\frac{\vec{p}\times\vec{a}(t,z)}{|\vec{a}(t,z)|}f_{lab}(t,z)$
encapsulates the kinetic-polarization effect.
The only terms in (\ref{eq:n10}) which couple the three components
of $\vec{\eta}_{lab}$ are the Thomas-BMT term and the 
Baier-Katkov correction term.

As mentioned above, there exists a system of
SDEs underlying (\ref{eq:n10}) (for details, see Ref. \citen{SDE}). 
In particular, $f_{lab}$ and $\vec{\eta}_{lab}$ are related to a
spin-orbit density ${\cal P}_{lab}={\cal P}_{lab}(t,z,\vec{s})$ via
\begin{eqnarray}
&& f_{lab}(t,z)=\int_{{\mathbb R}^3}\; d\vec{s}\;
{\cal P}_{lab}(t,z,\vec{s}) \; , 
\label{eq:3.20a} \\ 
&& \vec{\eta}_{lab}(t,z)=\int_{{\mathbb R}^3}\; d\vec{s}\;
\vec{s}\;{\cal P}_{lab}(t,z,\vec{s}) \; ,
\label{eq:3.21a}
\end{eqnarray}
where ${\cal P}_{lab}$ satisfies the Fokker-Planck equation corresponding to the
system of SDEs in Ref. \citen{SDE}. 
These SDEs can be used as the basis
for a Monte-Carlo spin tracking algorithm, i.e., for Method 2
mentioned in Section 1 above.
This would extend the standard Monte-Carlo spin tracking algorithms,
which we mentioned in Section 2 above,
by taking into account all physical effects described by (\ref{eq:n10}),
i.e., the Sokolov-Ternov effect,
the Baier-Katkov correction, the kinetic-polarization effect and, of course,
spin diffusion. 

If we ignore the spin-flip terms and the kinetic-polarization term 
in the BE then
(\ref{eq:n10}) simplifies to the RBE
%
\begin{eqnarray}
 \partial_t\vec{\eta}_{lab} =  L_{FP}^{lab}(t,z)\vec{\eta}_{lab}
+ \Omega^{lab}(t,z)\vec{\eta}_{lab} \; .
\label{eq:n11c}
\end{eqnarray}
The RBE models spin diffusion due to the effect of the
stochastic orbital motion on the spin and thus contains
those terms of the BE which are related to the
radiative depolarization rate $\tau_{dep}^{-1}$. This effect is clearly
seen in the SDEs (see, e.g., (\ref{eq:4.10}) and (\ref{eq:3.15})).
\section{The Reduced Bloch equation
in the beam frame}
In the beam frame, i.e., in the accelerator coordinates $y$ of Section 2,
the RBE (\ref{eq:n11c}) becomes
\ba
&& \partial_\theta\vec{\eta}
= L_{\rm FP}^y(\theta,y)\vec{\eta} + 
\Omega(\theta,y)\vec{\eta} \; .
\label{eq:3.10}
\ea
Because the coefficients of $L_{\rm FP}^y$ are $\theta$-dependent, the
RBE (\ref{eq:3.10}) is numerically and analytically quite complex.
So we first approximate it by treating the synchrotron radiation as a
perturbation. Then, in order to solve it numerically 
to determine the long-time behavior
that we need, we address the system of SDEs underlying (\ref{eq:3.10}) and 
apply the  refined averaging
technique presented 
in Ref. \citen{CT2} (see also \citen{El}),
for the orbital dynamics, and extend it to include spin.
The averaged SDEs are then used to construct an approximate
RBE which we call the effective RBE.

The system of SDEs underlying (\ref{eq:3.10})
reads as
\footnote{We denote the random dependent variables 
like $Y$ in (\ref{eq:4.10})
by capital letters to distinguish them from 
independent variables like $y$ in (\ref{eq:3.10}).}
\ba
&& \frac{dY}{d\theta}
= (A(\theta)+\epsilon_R \delta A(\theta))Y
+ \sqrt{\epsilon_R}\sqrt{\omega(\theta)}
e_6\xi(\theta) \; ,
\label{eq:4.10} \\
&& \frac{d\vec{S}}{d\theta} = 
[\Omega_0(\theta)+\epsilon_S C(\theta,Y)]\vec{S} \; ,
\label{eq:3.15}
\end{eqnarray}
where the orbital dynamics has been linearized in $Y$ and 
$\Omega=\Omega_0+\epsilon_S C$
has been linearized in $Y$ so that
\ba
&& C(\theta,Y) = \sum_{j=1}^6\; C_j(\theta)Y_j \; .
\label{eq:4.10a} 
\end{eqnarray}
Also, $A(\theta)$ is a Hamiltonian matrix representing the nonradiative
part of the orbital dynamics and
$Y$ has been scaled so that $\epsilon_R$ is the size of the orbital effect of
the synchrotron radiation. Thus $\epsilon_R \delta A(\theta)$ represents
the orbital damping effects due to
synchrotron radiation and the cavities,  
$\sqrt{\epsilon_R}\xi(\theta)$ represents the associated quantum
fluctuations, $\xi$ is the white noise process and $e_6:=(0,0,0,0,0,1)^T$.
In the spin equation (\ref{eq:3.15}), 
$\Omega_0$ is the closed-orbit contribution to $\Omega$ so that
$\epsilon_S C(\theta,Y)$ is what remains and $C(\theta,Y)$ is chosen
$O(1)$. Hence $\epsilon_S$ estimates the size of 
$\Omega-\Omega_0$. Both $\Omega_0(\theta)$ and $C(\theta,Y)$ are, of course, 
skew-symmetric $3\times 3$ matrices. We are interested in the
situation where $\epsilon_R$ and $\epsilon_S$ are small in some
appropriate sense.

Eqs. (\ref{eq:4.10}) and (\ref{eq:3.15}) can
%
%
be obtained by transforming the system of SDEs in Ref. \citen{SDE}
%
%
%
from the laboratory frame to the beam frame \cite{HBBE}.
However, since in this section we only deal with the RBE (not with the BE),
(\ref{eq:4.10}) and (\ref{eq:3.15}) 
can also be found in 
older expositions on spin in
high-energy electron storage rings, e.g., Ref. \citen{RR}.
Note that these expositions make approximations 
as for example with the linearity of (\ref{eq:4.10}) in $Y$
and the linearity of $C(\theta,Y)$ in $Y$. 

With (\ref{eq:4.10}) and (\ref{eq:3.15}) the evolution equation for
the spin-orbit joint probability density ${\cal P}={\cal P}(\theta,y,\vec{s})$
is the following spin-orbit Fokker-Planck equation
\ba
&& \partial_\theta {\cal P} = L_{\rm FP}^y(\theta,y){\cal P}
-\partial_{\vec{s}}\cdot
\Biggl(\biggl(\Omega(\theta,y)\vec{s}\biggr)
{\cal P}\Biggr) \; ,
\label{eq:3.20}
\end{eqnarray}
where $L_{\rm FP}^y$
is the orbital Fokker-Planck operator.
The phase-space density $f$ and the
polarization density $\vec{\eta}$ corresponding
to ${\cal P}$ are defined by
\begin{eqnarray}
&& f(\theta,y) = \int_{{\mathbb R}^3}\; d\vec{s}\;
{\cal P}(\theta,y,\vec{s}) \; , \quad
\vec{\eta}(\theta,y)=
\int_{{\mathbb R}^3}\; d\vec{s} \;
\vec{s}\;{\cal P}(\theta,y,\vec{s}) \; ,
\label{eq:3.22}
\end{eqnarray}
which are the beam-frame analogs of (\ref{eq:3.20a}) and 
(\ref{eq:3.21a}).
The local polarization vector $\vec{P}_{loc}$ from Section 2 above is
related to $f$ and $\vec{\eta}$ by
\begin{eqnarray}
&& \vec{\eta}(\theta,y) = f(\theta,y)\vec{P}_{loc}(\theta,y)
\; .
\label{eq:2.12}
\end{eqnarray}
The RBE (\ref{eq:3.10}) follows from
(\ref{eq:3.20}) by differentiating (\ref{eq:3.22}) w.r.t. $\theta$
and by using the Fokker-Planck equation for ${\cal P}$.
This proves that (\ref{eq:4.10}) and (\ref{eq:3.15}) is the system of SDEs
which underlie the RBE (\ref{eq:3.10}).
For (\ref{eq:3.10}), see also Ref. \citen{BH}.
\section{The Effective Reduced Bloch equation
in the beam frame}
The effective RBE is, by definition, an approximation of the RBE 
(\ref{eq:3.10}) obtained by approximating the system of SDEs
(\ref{eq:4.10}) and (\ref{eq:3.15})  
using the method of averaging, see Refs. \citen{El}-\citen{CE}.
We call the system of SDEs
underlying the effective RBE  the effective system of SDEs.
We now discuss first-order averaging in the case where
$\epsilon:=\epsilon_S=\epsilon_R$ is small.

To apply the method of averaging to (\ref{eq:4.10}) and (\ref{eq:3.15})
we must transform them to a standard form for averaging, i.e., we must
transform the variables $Y,\vec{S}$ to slowly varying variables.
We do this by using a fundamental solution matrix $X$ of the
unperturbed $\epsilon=0$ part of (\ref{eq:4.10}), i.e.,
%
\ba
&& X'=A(\theta)X \; ,
\label{eq:4.18}
\end{eqnarray}
and a fundamental solution matrix $\Phi$ of the
unperturbed $\epsilon=0$ part of (\ref{eq:3.15}), i.e.,
\ba
&& \Phi'=\Omega_0(\theta)\Phi\; .
\label{eq:4.18new}
\end{eqnarray}
%
We thus transform $Y$ and $\vec{S}$ into the slowly varying
$U$ and $\vec{T}$ via
\ba
&& Y(\theta)=X(\theta) U(\theta) \; , \quad  
\vec{S}(\theta)=\Phi(\theta)\vec{T}(\theta) \; .
\label{eq:4.20}
\end{eqnarray}
Hence (\ref{eq:4.10}) and (\ref{eq:3.15}) are transformed to
\ba
&& U' = \epsilon {\cal D}(\theta)U
+ \sqrt{\epsilon}\sqrt{\omega(\theta)}
X^{-1}(\theta)e_6\xi(\theta) \; ,
\label{eq:4.31} \\
&& \vec{T}'=\epsilon{\mathfrak D}(\theta,U)\vec{T} \; ,
\label{eq:4.25} 
\end{eqnarray}
where ${\cal D}$ and ${\mathfrak D}$
are defined by
\ba
&& {\cal D}(\theta):=X^{-1}(\theta)\delta A(\theta)X(\theta) \; ,
\label{eq:4.28} \\
&& {\mathfrak D}(\theta,U)
:=\Phi^{-1}(\theta)C(\theta,X(\theta)U)\Phi(\theta) \; .
\label{eq:4.26}
\end{eqnarray}
Of course, (\ref{eq:4.31}) and (\ref{eq:4.25}) carry the same information 
as (\ref{eq:4.10}) and (\ref{eq:3.15}).
Now, applying the method of averaging to (\ref{eq:4.31}) and (\ref{eq:4.25}), 
we obtain the following effective system of SDEs
\ba
&& V' = \epsilon \bar{\cal D}V + \sqrt{\epsilon}{\cal B}
(\xi_1,...,\xi_k)^T \; ,
\label{eq:4.40}\\
&& \vec{T}_a'=\epsilon\bar{\mathfrak D}(V)\vec{T}_a \; ,
\label{eq:3.15e}
\end{eqnarray}
where the bar denotes $\theta$-averaging, i.e., the operation
$\lim_{T\rightarrow\infty}(1/T)\int_0^T d\theta\cdots$.
Moreover $\xi_1,...,\xi_k$ are statistically independent versions of
the white noise process and
${\cal B}$ is a $6\times k$ matrix which satisfies
${\cal B}{\cal B}^T=\bar{\cal E}$
with $k=rank(\bar{\cal E})$ and where $\bar{\cal E}$
is the $\theta$-average of
\ba
&& {\cal E}(\theta)
= \omega(\theta)X^{-1}(\theta)e_6^{} e_6^TX^{-T}(\theta) \; .
\label{eq:4.29}
\end{eqnarray}
For physically reasonable $A$ and $\Omega$
the fundamental matrices $X$ and $\Phi$ 
are quasiperiodic functions whence
${\cal D},{\mathfrak D}(\cdot,U)$ and ${\cal E}$ 
are quasiperiodic functions so that
their $\theta$  averages $\bar{\cal D},\bar{\mathfrak D}(V)$ 
and $\bar{\cal E}$ exist.

Our derivation of (\ref{eq:4.40}) from (\ref{eq:4.31}) is discussed
in some detail in Ref. \citen{SDE}. We are close to showing
that $U=V+O(\epsilon)$ on $\theta$-intervals of length $O(1/\epsilon)$ 
and it seems likely that this error is valid for $0\leq \theta<\infty$,
because of the radiation damping. This is a refinement 
of Ref. \citen{CT2}
and assumes a non-resonance condition. 
Since the sample paths of
$U$ are continuous and $U$ is slowly varying it seems likely that 
$\vec{T}_a$ is a good approximation to $\vec{T}$ and we are working
on the error analysis. Spin-orbit resonances will be an
important focus in the construction of 
$\bar{\mathfrak D}(V)$ from (\ref{eq:4.26}) which contains both the
orbital frequencies in $X$ and the spin precession frequency in $\Phi$.

Since, by definition, the effective system of SDEs 
underly the effective RBE, the latter can be obtained from the former
in the same way as we obtained (\ref{eq:3.10}) from 
(\ref{eq:4.10}) and (\ref{eq:3.15}) (recall the discussion after (\ref{eq:3.22})).
Thus the evolution equation for
the spin-orbit probability density
${\cal P}_{V}={\cal P}_{V}(\theta,{\rm v},\vec{t})$
is the following Fokker-Planck equation:
\ba
&&  \hspace{-10mm}
\partial_\theta {\cal P}_{V} = L^V_{\rm FP}({\rm v}){\cal P}_{V}
-\epsilon\partial_{\vec{t}}\cdot
\Biggl(\biggl( \bar{\mathfrak D}({\rm v})\vec{t}\biggr)
{\cal P}_{V}\Biggr) \; ,
\label{eq:4.50}
\end{eqnarray}
where
\ba
&& L^V_{\rm FP}({\rm v})
= -\epsilon\sum_{j=1}^6 \partial_{{\rm v}_j}
(\bar{\cal D} {\rm v})_j
+ \frac{\epsilon}{2}\sum_{i,j=1}^6 \bar{\cal E}_{ij}
\partial_{{\rm v}_i}\partial_{{\rm v}_j} \; .
\label{eq:4.51}
\end{eqnarray}
The polarization density $\vec{\eta}_V$ corresponding
to ${\cal P}_{V}$ is defined by
\begin{eqnarray}
\vec{\eta}_V(\theta,{\rm {\rm v}})
=\int_{{\mathbb R}^3}\; d\vec{t}\;
\vec{t}\;{\cal P}_{V}(\theta,{\rm v},\vec{t}) \; ,
\label{eq:4.52}
\end{eqnarray}
so that,  by (\ref{eq:4.50}), the effective RBE is
\ba
&& \partial_\theta\vec{\eta}_V 
= L^V_{\rm FP}({\rm v})\vec{\eta}_V + 
\epsilon\bar{\mathfrak D}({\rm v})\vec{\eta}_V \; .
\label{eq:4.53}
\ea
This then is the focus of our approach in Method 1.
For more details on this section, 
see Refs. \citen{SDE}, \citen{HABBE} and \citen{OB}. 

\section{Next steps}

\begin{itemize}
\item Further development of Bloch-equation approach (numerical and 
theoretical), i.e., of Method 1 and with a realistic lattice.
\item Development of validation methods, i.e., Methods 2-4.
Note that Method 2 is an extension of
the standard Monte-Carlo spin tracking algorithms and for that matter
we will study Refs. \citen{KP}, \citen{KPS} and \citen{PB}.
\item Investigating the connection between the Bloch-equation approach and
the standard approach based on  the
Derbenev-Kondratenko formulas, 
and studying the potential for correction terms \cite{DBBQ}
to $\tau_{\rm DK}^{-1}$ by using the RBE.
\end{itemize}

\section{ Acknowledgement}
This material is based upon work supported by the U.S. Department
of Energy, Office of Science, Office of High Energy Physics, under Award
Number DE-SC0018008.


\end{document}